\documentclass[useAMS,fleqn,usenatbib]{mnras}

\usepackage[export]{adjustbox}
\usepackage{txfonts}
\usepackage{tikz}
\usepackage{hyperref}

\definecolor{lime}{HTML}{A6CE39}
\DeclareRobustCommand{\orcidicon}{%
    \begin{tikzpicture}
    \draw[lime, fill=lime] (0,0) 
    circle [radius=0.16] 
    node[white] {{\fontfamily{qag}\selectfont \tiny ID}};
    \draw[white, fill=white] (-0.0625,0.095) 
    circle [radius=0.007];
    \end{tikzpicture}
    \hspace{-2mm}
}

\newcommand{\orcid}[1]{\href{https://orcid.org/#1}{\orcidicon}}

\def\R200 {R_{200}}

\title[Cosmic Kite]{Cosmic Kite: Auto-encoding the Cosmic Microwave Background.}

\author[de los Rios et al.]{\parbox[t]{\textwidth}{
Mart\'in de los Rios,$^{1,2,3}$\thanks{E-mail: martindelosrios13@gmail.com}\orcid{0000-0003-2190-2196}}
\\
\\
$^{1}$Instituto de F\'isica Te\'orica, UAM-CSIC, c/ Nicol\'as Cabrera 13-15, Universidad Aut\'onoma de Madrid, Cantoblanco, Madrid 28049, Spain\\
$^{2}$Departamento de F\'isica Te\'orica, Universidad Aut\'onoma de Madrid, 28049 Madrid, Spain\\
$^{3}$ICTP South American Institute for Fundamental Research \& Instituto de F\'isica Te\'orica, Universidade Estadual \\ Paulista, 01140-070, S\~ao Paulo-SP, Brazil\\
\\ 
}

\begin{document}

\date{Accepted . Received ; in original form }

\pagerange{\pageref{firstpage}--\pageref{lastpage}} \pubyear{2020}

\maketitle

\label{firstpage}

\begin{abstract}

In this work we present the results of the study of the cosmic microwave background TT power spectrum through auto-encoders
in which the latent variables are the cosmological parameters. 
This method was trained and calibrated using a data-set composed by $80000$ power spectra from random cosmologies computed numerically with the \texttt{CAMB} code.
Due to the specific architecture of the auto-encoder, the encoder part is a model that estimates the maximum-likelihood
parameters from a given power spectrum.
On the other hand, the decoder part is a model that computes the power spectrum from the cosmological
parameters and can be used as a forward model in a fully Bayesian analysis.
We show that the encoder is able to estimate the true cosmological parameters
with a precision varying from $\approx 0.004 \%$ to $\approx 0.2 \%$ (depending on the cosmological parameter),
while the decoder computes the power spectra with a mean percentage error of $\approx 0.0018 \%$ for all the multipole range.
We also demonstrate that the decoder recovers the expected trends when varying the cosmological parameters one by one, and that it does not introduce any significant bias on the estimation of cosmological parameters through a Bayesian analysis.
These studies gave place to the \texttt{Cosmic Kite} python software that is publicly available and can be downloaded and installed from \href{https://github.com/Martindelosrios/cosmic-kite}{https://github.com/Martindelosrios/cosmic-kite}.
Although this algorithm does not improve the precision of the measurements compared with the traditional methods, it reduces
significantly the computation time and represents the first attempt towards forcing the latent variables to have a physical interpretation.
\end{abstract}


\section{Introduction}

Machine learning techniques represent a new 
way of analysing big data-sets in an agnostic and homogeneous way. Taking into
account the amount of data generated by current and future surveys and 
simulations, the data-driven techniques will become a fundamental tool for their analysis. 
These methods are very useful and powerful tools to find 
patterns and relations between the variables that are involved in a specific problem. 
These kind of data-driven techniques have been applied with a lot of success in several astronomical problems in the last decade.
In particular there are several works which analyse the Cosmic Microwave Background (CMB) at map level.
From these, we can mention,  \cite{montefalcone} where they reconstruct the masked regions of the CMB with neural networks,
\cite{khatri} where they applied
data-driven methods to separate the different components of the CMB, \cite{norgaard} where they
performed a foreground removal from the polarization map with neural networks and
\cite{deepsphere} where they adapt the traditional layers used in convolutional neural networks 
to work within the \texttt{HEALPix} scheme \cite{healpix} guaranteeing  spherical simmetry.

It is also important to point out the codes \texttt{PICO} \citep{pico}, \texttt{COSMONET} \citep{cosmonet1, cosmonet2}
and \texttt{COSMO-POWER}\footnote{The \texttt{COSMO-POWER} code was released while this work was under review.}\citep{cosmopower} that have the same spirit of our method.
These codes provides fast emulators of the CMB power spectra that can be used as forward models in an MCMC analysis. 
Nevertheless, as it will be explained in the following sections, our approach, due to the specific auto-encoder architecture, provides an
end-to-end analysis of the power spectrum that then can be decomposed on two models.
One model is a fast emulator of the power spectrum, similar to the mentioned codes,
that can be used as forward model in MCMC analysis. 
While the second model can be used to predict the cosmological parameters that
best fits a given power spectrum without the need of performing an MCMC analysis.

The CMB is one of the most important and well studied observations of the early universe \citep{planck2018}. 
Due to its capacity to impose constraints on some cosmological parameters, it is consider one of the most relevant probes supporting the standard cosmological model.

The traditional way of estimating the cosmological parameters using the CMB is by measuring the power spectrum of the temperature fluctuations and comparing it with the theoretical predictions computed by solving numerically the Boltzmann-Einstein equations. 

In general, this comparison is made in a Bayesian fashion through an extensive exploration in the
cosmological parameter space using Monte Carlo Markov chains (MCMC). This method has the advantage that it does not only estimate the best-fit parameters, but also give us information about the full posterior of these parameters around the best fit values. 
However, the main disadvantage of these algorithms is that they consume a lot of time, especially when they want to explore highly dimensional spaces such as those of cosmological interest in which is common to work with around $ 6 $ variables.

In this work, we analyse the temperature-temperature (TT) power spectrum of the CMB using auto-encoders architectures.
Auto-encoders, as it will be explained in section \ref{sec:metodos_cmb}, are an specific type of neural network 
\citep{autoencoder, autoencoders} (neural networks with a bottle-neck layer) that are specially good for finding the
most relevant features of a given data-set, and, hence, can be used for dimensionality reduction.
One of the main disadvantage of them is that, as is common to all the highly non-linear methods,
it is difficult to obtain a physical interpretation of their results. 
This, in turn, impose a restriction on the applicability of such methods on physical problems, and hence is one of the hottest topics inside the current machine learning investigations.
In order to obtain a physical interpretation of the results we impose a custom loss function to the latent variables that force them to be the cosmological parameters.
The benefit of this analysis is two-fold. On one hand, the encoder represents a function that gives the cosmological parameters that best fits a given power spectrum.
On the other hand, the decoder represents a function that gives the TT power spectrum that correspond to a given cosmological model.
This function can be used as a forward model in Bayesian analysis, replacing the standard codes that solve numerically the relevant equations.
It is worth remarking that, although the main results of the paper were
obtained using only the TT power spectrum, given the proper training-set all the model can be easily
extended to the TE and EE power spectra as shown in appendix \ref{ap:TE-EE} or to any other
quantity related to the cosmological parameters.

This paper is organized as follows. The data that we will use in all our analysis is presented in section \ref{sec:set_datos}. 
In section \ref{sec:metodos_cmb} we present the main auto-encoder analysis of the CMB power spectra that includes the estimation of the encoder performance in predicting the
best-fit cosmological parameters \ref{sec:encoder} and the quantification of the errors introduced by the decoder on the estimation of the theoretical power spectra \ref{sec:decoder} and on the full Bayesian analysis \ref{sec:bayes}. 
In section \ref{sec:cosmickite} we analyze the main features of the \texttt{cosmic-kite} software, including examples of basic usage and profiling.
Finally in section \ref{sec:conclusions}, we present a few concluding remarks.

\section{Construction of the data set.} \label{sec:set_datos}

The current standard cosmological model has been established very robustly over the last decades, achieving a very high precision in the
cosmological parameters measurements. 
Although there are some discrepancies \citep{riess}, the most recently data favors an standard cosmology
with $\Omega_{c}h^{2}= 0.120 \pm 0.0012$, $\Omega_{b}h^{2}=0.02237 \pm 0.00015$, $\Omega_{\Lambda}=0.6847 \pm 0.0073$, $H_{0} = 67.36 \pm 0.54$, $n=0.9649 \pm 0.0042$, $A_{s}=(2.098 \pm 0.101)10^{-9} $ and $\tau = 0.0544 \pm 0.0073$ \citep{planck2018}, implying that we are living in a flat universe currently dominated by dark energy.

This concordance model is fully established on the assumption that the universe is statistically homogeneous and isotropic, and that can be
model with tiny perturbations over a background FRWL metric \citep{dodelson}.
Taking into account the high densities found in the early universe before recombination, it is mandatory to analyze the behaviour and interactions of the different components. 
The temperature fluctuations of the Cosmic Microwave Background are the result of the evolution of the different components in the early Universe and the statistics of such physical phenomena can be explained through the Boltzmann-Einstein equations. 
There are several codes that solves the set of coupled Boltzmann-Einstein equations numerically and have been extensively used, such as \texttt{CAMB (Code for Anisotropies in the Microwave Background)} \citep{camb} \footnote{\href{https://camb.info/}{https://camb.info/}} and \texttt{CLASS} (Cosmic Linear Anisotropy Solving System) \citep{class1, class2}. 
These codes have the advantage of solving the full set of coupled Boltzmann-Einstein equations (up to some order of precision depending on the code), but they can be very computationally expensive.
In this work we will make an extensive use of the  code \texttt{CAMB} code for the creation of the data-set. 

In order to build our data-set, we estimate the CMB TT power spectra,
using the \texttt{CAMB} code with the default \verb|planck_2018.ini| file, for $ 80000 $ random cosmological models varying the cosmological parameters
$\Omega_{c}h^{2}$, $\Omega_{b}h^{2}$, $H_{0}$, $n$, $A_{s}$ and $\tau$
within an interval around the parameters estimated by \cite{planck2018}.
In table \ref{table:params} we present the minimum and maximum cosmological parameter values for the intervals used to build the data-set.

\begin{table}
\centering
\begin{tabular}{cccc} 
 Parameter & Minimum & Maximum & Planck \\
 \hline
 $\Omega_{c}h^{2}$ & $0.1096$  & $0.130 $  & $0.120$ \\
 $\Omega_{b}h^{2}$ & $0.02128$ & $0.02348$ & $0.02237$ \\
 $H_{0}$           & $58.12$   & $76.52$   & $67.36$ \\
 $n$               & $0.9375$  & $0.9945$  & $0.9649$ \\
 $A_{s}$           & $1.930*10^{-9}$ & $2.270*10^{-9}$ & $2.098*10^{-9}$ \\
 $\tau$            & $0.014$   & $0.094 $  & $0.0544$ \\
\end{tabular}
\caption{Minimum and maximum cosmological parameter values used to build the dataset.} \label{table:params}
\end{table}

It is important to note that the spectra computed by numerically solving the Boltzmann-Einstein equation are theoretical, 
and hence are not affected by any other contamination source.
On the other hand, the spectra estimated from the observational temperature
fluctuation suffers from various observational problems such as, contamination from secondary anisotropies \citep{secondaries_ans} point sources,
mask effects, resolutions effects, etc \citep{hivon}.
Each one of these effects introduce errors in the measurement that must be taken into account before any comparison with a theoretical model.
As there are several methods and codes to mitigate the effect of these problems \citep{hivon,namaster}, we will focus on analysing theoretical spectra and leave for another work the possibility of mitigating these problems with machine learning techniques.

\begin{figure}
 \centering
 \includegraphics[scale=0.6]{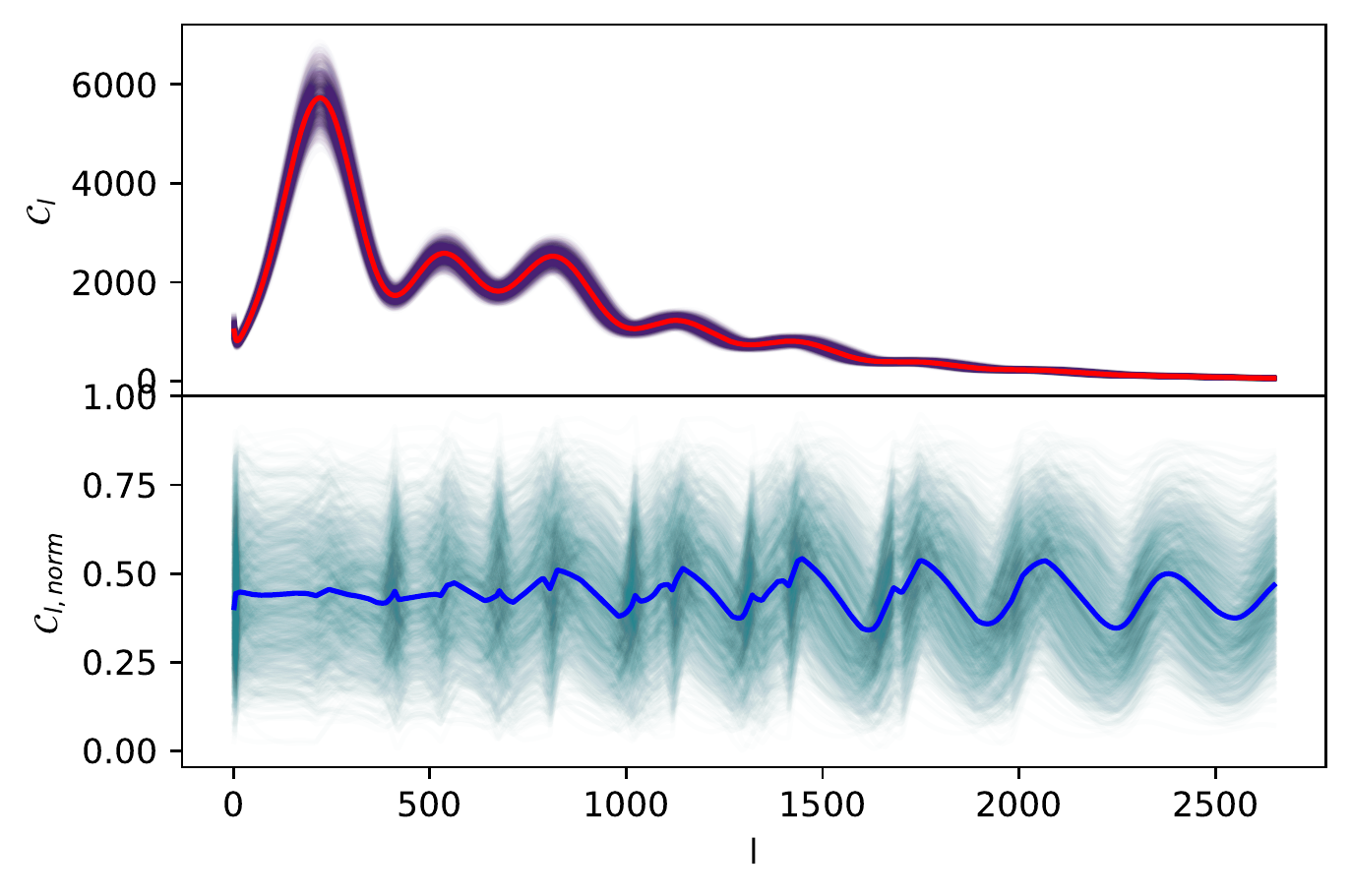}
 \caption{\textit{Upper Panel:} Examples of theoretical power spectra computed with the \texttt{CAMB} code. \textit{Lower Panel:} Examples of theoretical power spectra computed with the \texttt{CAMB} code and pre-processed with a minmax scaler.}
 \label{fig:espectros_bin}
\end{figure}

In the top-panel of figure \ref{fig:espectros_bin} we show examples of theoretical power spectra that belong to the dataset computed with \texttt{CAMB}.

In order to avoid over-fitting and to have a robust measure of the performance of the method
we split our data-set in $3$ mutually independent subsets:
a training-set (consisting on the 75\% of the full sample), 
a validation-set (consisting on the 20\% of the full sample) and a testing-set (consisting on the 5\% of the full sample). The training-set will be used for training the model, the validation-set will be used to measure the performance at the end of each training epoch, and, finally, the testing-set will be used for measuring the performance of the method after the full training.

As is common in machine learning techniques, in order to improve the
performance of the methods, we normalize the power spectra and the cosmological parameters in order
to have all our data between 0 and 1. 
In order to do this, we implement a \verb|minmax scaler| from the \texttt{python} package \texttt{scikit-learn} \citep{scikit-learn} calibrated with the training-set.
In the bottom-panel of figure \ref{fig:espectros_bin} we show examples of theoretical power spectra from our dataset that were computed with the \texttt{CAMB} code and normalized within 0 and 1.

\section{\texttt{Cosmic Kite}: Auto-encoding the Cosmic Microwave Background.} \label{sec:metodos_cmb}

A neural network is a mathematical model inspired 
by the structure of biological brains. 
The main purpose of a such model is to construct an approximate function that associates input $x$ data
with output data $y=f(x)$. 
This kind of models generally consists of an input layer, some hidden layers, and an output layer. 
In turns, each layer consists on several neurons.
Each neuron (also called nodes) is represented by a non-linear function that takes as input the weighted sum of the outputs of the previous layer.
These weights are tunned, during the training process, to minimize a given loss function.
Therefore, the final function will be a composition of several non-linear functions that correspond to the nodes of each layer.

\begin{figure}
 \includegraphics[scale=0.34]{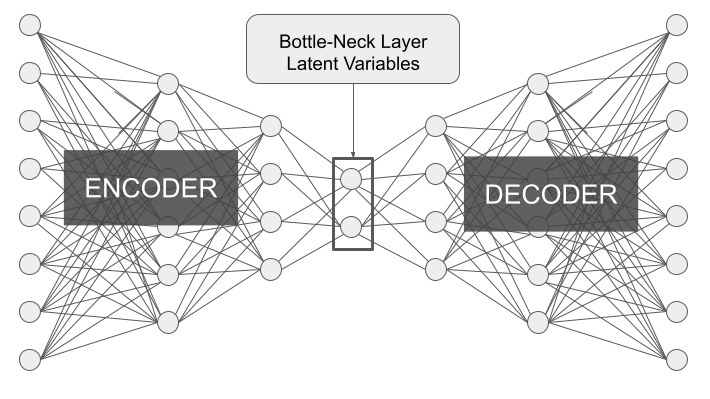}
 \caption{Schematic figure for representing the Auto-encoder architecture. The encoder model is
 composed by all the layers before the bottle-neck layer while the decoder model is composed by all the 
 layers after the bottle-neck layer.}
 \label{fig:autoencoder}
\end{figure}

Auto-encoders are neural networks designed to compress data and pick out the relevant features of a given data-set \citep{autoencoder}.
They are characterized by their specific architecture composed by a fully-connected neural networks with a bottle-neck layer in the middle i.e, a hidden layer with few neurons (see figure \ref{fig:autoencoder} for an schematic representation).
Also, in most of the implementations, the input and the output are the same and with a much higher dimension than the bottle-neck layer.
As the auto-encoder is force to learn exactly the input data passing through a low-dimensional layer (the bottle-neck hidden layer), the data is compressed into the latent variables (neurons of the bottle-neck layer), and then decoded back to the original form.
It is common to call \textit{encoder} to the neural network made up by the layers before the bottle-neck layer, and \textit{decoder} to the neural network formed by the layers after the bottle-neck layer.
It is worth remarking that in this work we extensively used the python packages \texttt{tensorflow} \citep{tensorflow} and \texttt{keras}\citep{keras} for the implementation of the neural networks.
In our work, we used a fully connected auto-encoder with $7$ hidden layers with $1000$, $100$, $50$, $6$,
$50$, $100$ and $1000$ neurons respectively, where the
input and output layers correspond to the normalized
power spectra which multipoles from $l=2$ to $l=2650$.
We tested different activation functions and the best results were obtained when the neurons are activated with a \texttt{LeakyRelu} activation function with $\alpha = 0.1$.
Taking into
account the architecture, the auto-encoder learn to
compress the full power spectra into $6$ latent variables that correspond to the neurons of our bottle-neck layer.

For both the decoder and the encoder we choose a mean squared error (MSE) loss function shown in equations \ref{eq:dec_loss} and \ref{eq:enc_loss}.

\begin{eqnarray} \label{eq:dec_loss}
    DEC_{loss} &=& \frac{1}{2648} \sum_{i=2}^{2650} \left( C_{l}^{R} - C_{l}^{P} \right)^{2} \\
    ENC_{loss} &=& \frac{1}{6} \sum_{i=1}^{6} \left( \Omega_{i} - \mathcal{L}_{i}  \right)^{2} \label{eq:enc_loss}
\end{eqnarray}

where in eq. \ref{eq:dec_loss} $ C_{l}^{R} $ and $ C_{l}^{P} $ are the real and predicted values of each multipole of the power spectra normalized between $1$ and $0$
and in eq. \ref{eq:enc_loss}  $\Omega_{i}^{R}$ are the real cosmological parameters normalized between 1 and 0 used  to compute each power spectra, and $\mathcal{L}_{i}$ are the $6$ latent variables.
It is worth to recap that, although the CMB temperature distribution is Gaussian, the $\mathcal{C}_{l}$ distribution is not.
Nevertheless, for high-l the central-limit theorem ensures that the
likelihood is well represented by a Gaussian.
Although, strictly speaking the likelihood for low-l is not Gaussian, we found that including or excluding the low multipoles
does not make any difference on the final predictions.
For this reason, and looking for a code that predict
all the multipoles at once, we decided to keep working with the full multipole range.

These loss functions guarantees that the latent variables will predict the real cosmological parameters and that the predicted power spectra will converge to the real ones.

Finally we define the loss function for the full auto-encoder as a weighted sum of the decoder and encoder losses.

\begin{equation}
    LOSS = \frac{1}{N}\sum_{j=1}^{N} \left[ \omega_{enc}ENC_{loss,j} + \omega_{dec}DEC_{loss,j} \right]
\end{equation}

where the summation runs over all the samples of the training-set.
It also important to remark that the weights $\omega_{enc}$ and $\omega_{dec}$ can be tuned in order to find the best performance or
to focus the training to give more importance to one of the losses.
After trying different values (we vary each weight from $10^{-5}$ to $10^{3}$) for the weights we found that the best
performance correspond to $\omega_{enc} = 1$ and $\omega_{dec}=10^{2}$. 
These weights do not favour neither loss, but just normalized them to have the same dynamical range.

\begin{figure}
    \centering
    \includegraphics[scale=0.6]{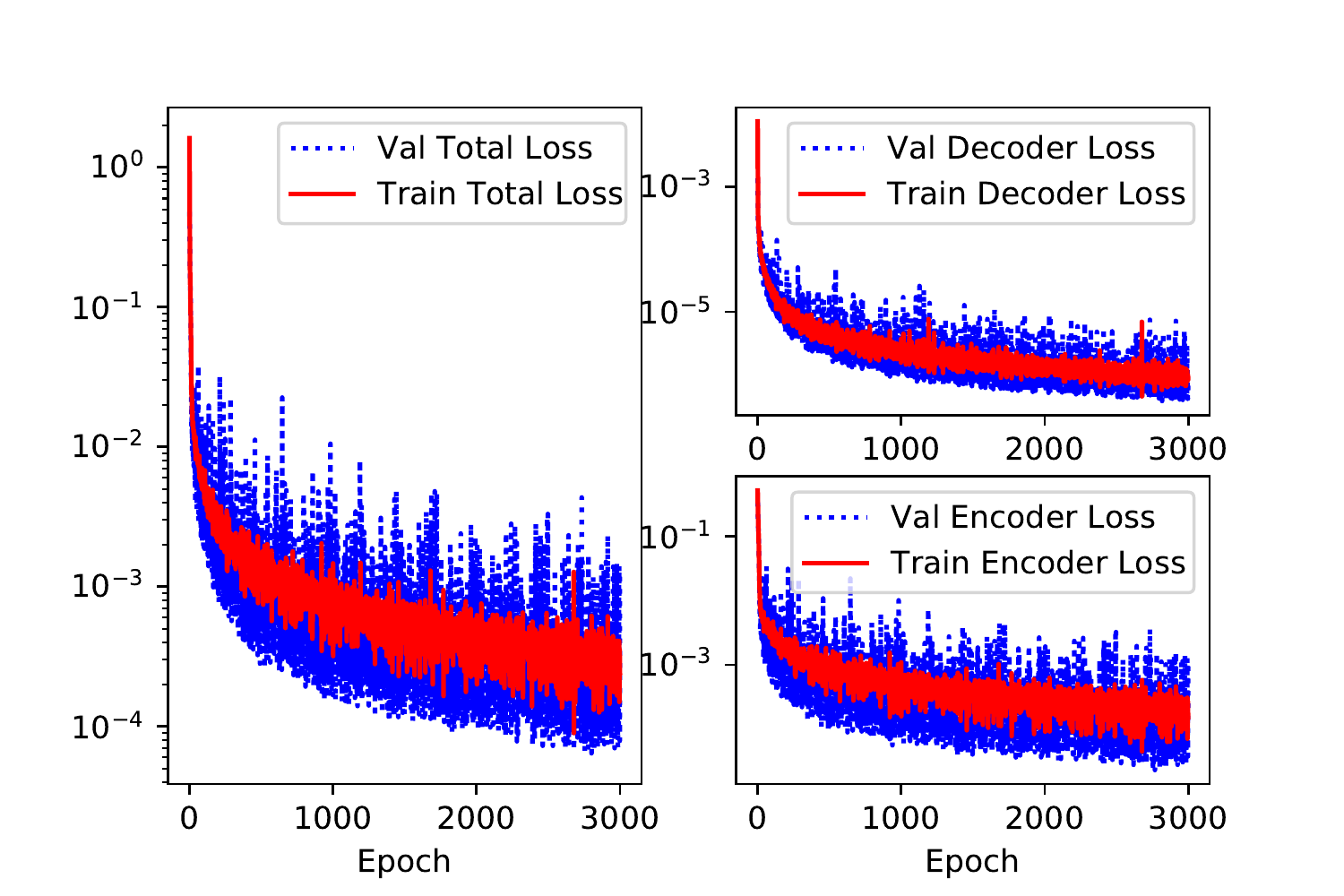}
    \caption{Training and validation loss obtained with a learning rate of $10^{-4}$ and a batch size of $256$. 
    In the left panel is shown the total training (blue lines) and validation (red dotted lines) loss. While in the right panels are shown the decoder (upper panel) and encoder (lower panel) training and validation loss with the same colour coding.
    }
    \label{fig:loss}
\end{figure}

After training with different learning rates (varying from $10^{-5}$ to $10^{-1}$) and different batch sizes (varying from $16$ to $512$), we found that the best
performance was achieved by training the auto-encoder for $3000$ epochs with an Adam optimizer \citep{adam} with a learning rate of $10^{-4}$ and a batch-size of $256$.  We show in the left panel of figure \ref{fig:loss}, the total training and validation loss. While in the right panel of figure \ref{fig:loss} are shown the decoder (upper panel) and 
encoder (lower panel) training and validation loss.

It can be seen how in the first epochs the networks minimize both the decoder and encoder
loss (and hence the full loss) in both the validation and training sets, until after $\approx 2000$ epochs the validation losses converges to a minimum.
The fact that the encoder validation loss have been minimized proves that the network is learning the relation between
the TT power spectra (input variables) and the
cosmological parameters (latent variables, i.e output variables of the encoder).

On the other hand, the fact that the decoder validation loss have been minimized proves
that the network is able to estimate
the TT power spectra (output variables) from the cosmological parameters (latent variables, i.e input variables of the decoder).

\subsection{Encoding: Estimating the best-fit parameters of the CMB.} \label{sec:encoder}

After we trained our auto-encoder, we applied it to the power spectra of the testing set. 
We remind that this dataset was not used neither for the training nor the validation. 
As we already mention our model has two output. On one hand the
encoder has as an output $6$ latent variables that will correspond to the cosmological parameters, and so, will be called "predicted cosmological parameters" $\Omega_{i}^{P}$ for now on.
On the other hand, the decoder has as an output the same power spectra that we put as an input but after the reconstruction from the latent variables.

\begin{figure*}
    \centering
    \includegraphics[scale = 0.6]{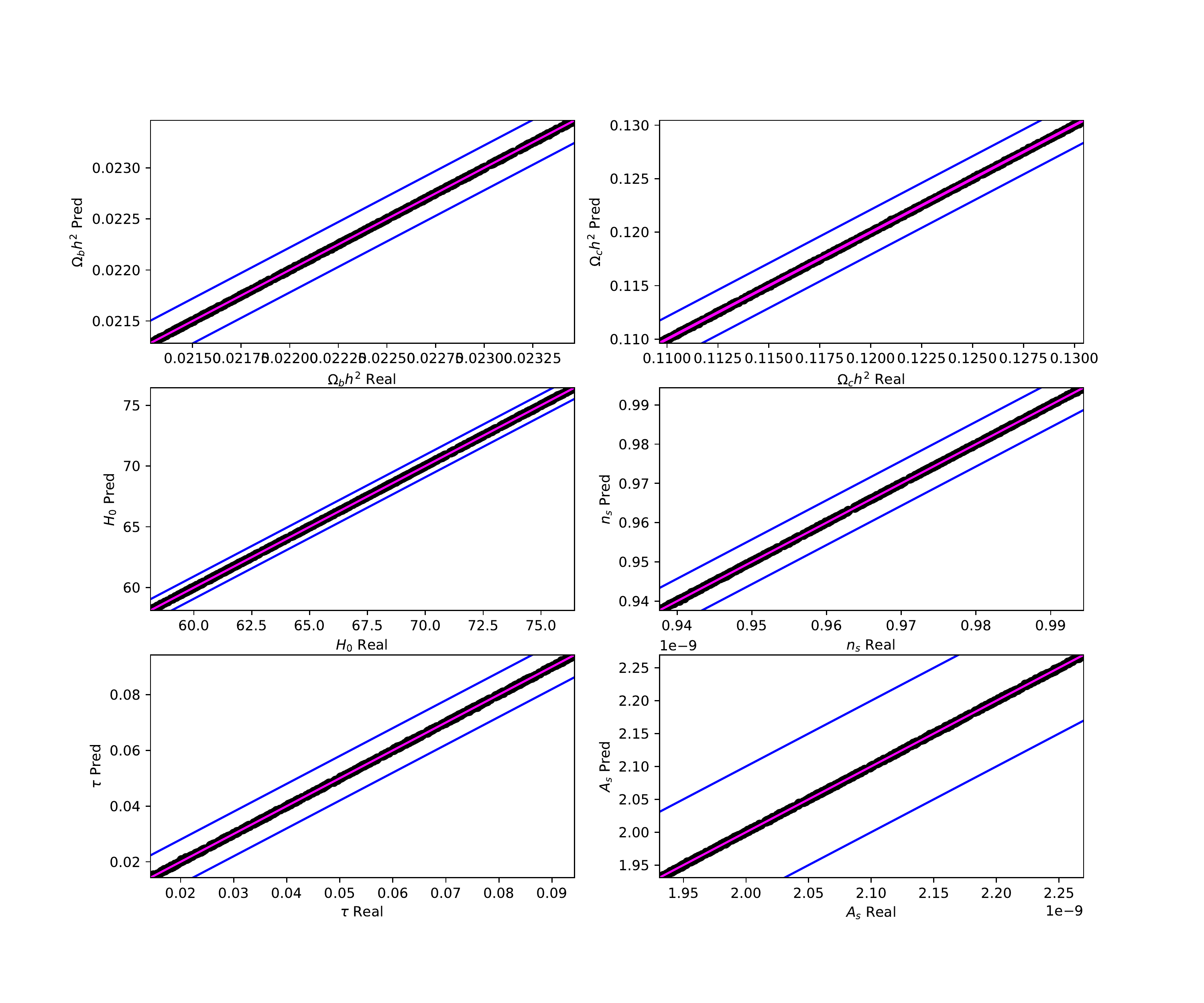}
    \caption{Real vs predicted cosmological parameters. The predicted parameters correspond
    to the latent variables that are the output of the encoder model. In blue lines are shown the errors in the Bayesian estimation reported by the Planck Collabortion \citep{planck2018}.}
    \label{fig:predicciones_cmb}
\end{figure*}

In figure \ref{fig:predicciones_cmb} we show the true cosmological parameters vs the predicted ones obtained by the encoder.
As we can see the encoder is capable of predicting with a very high accuracy all the cosmological parameters.

In order to quantify the accuracy reached by the encoder we estimate the mean-squared-error as:

\begin{equation}
    \chi^{2}_{i} = \;  < (\Omega_{i}^{R} - \Omega_{i}^{p})^{2} >
\end{equation}

and the precision as:

\begin{equation}
    prec_{i} = \;  < 100*\frac{|\Omega_{i}^{R} - \Omega_{i}^{p}|}{\Omega_{i}^{R}} >
\end{equation}

where the index $i$ runs over the $6$ cosmological parameters and the average is taken over all the testing-set.

The results of the statistics are summarised in table \ref{table:results}, and proves that
the good performance of the encoder in estimating the cosmological
parameters from a given TT power spectrum. 

\begin{table}
\begin{small}
\begin{tabular}{c|c|c|c|c|c|c}
 &  $\Omega_{b}h^{2}$ & $\Omega_{c}h^{2}$ & $H_{0}$ & $n$ & $\tau$ & $A_{s}$  \\ 
\hline
$\chi$  & $2.2e^{-6}$ & $2.6e^{-5}$ & $0.01$ & $5.4e^{-5}$ & $1.0e^{-4}$ & $5.1e^{-13}$  \\
\hline
$prec$  & $0.007$ & $0.017$ & $0.016$ & $0.004$ & $0.2$ & $0.019$  \\
\hline
$100(\sigma/\Omega)_{Pl}$  & $0.067$ & $1$ & $0.8$ & $0.43$ & $13.4$ & $4.83$  \\
\end{tabular}
\caption{Results of the performance of the encoder model in predicting the cosmological parameters. For a better comparison we add, in the last row, the percentage relative error obtained by the Planck collaboration in the estimation of each cosmological parameter.} \label{table:results}
\end{small}
\end{table}

It is worth remarking that
the encoder by itself is not able to estimate the full posterior probability of the parameters.
This model is only predicting the cosmological parameters that were used to compute
numerically the corresponding power spectra. This parameters, in turn, correspond to the maximum-likelihood parameters in traditional Bayesian analysis.

\subsection{Decoding: Estimating the CMB power spectra.}\label{sec:decoder}

As mentioned before, the decoding part computes the power spectra from the latent variables, i.e from the cosmological parameters.
Applying the trained decoder model to the
cosmological parameters used to build the testing-set,
we were able to compute the corresponding power spectra and compare it with the real ones computed with the \texttt{CAMB} code.

\begin{figure}
    \centering
    \includegraphics[scale=0.6]{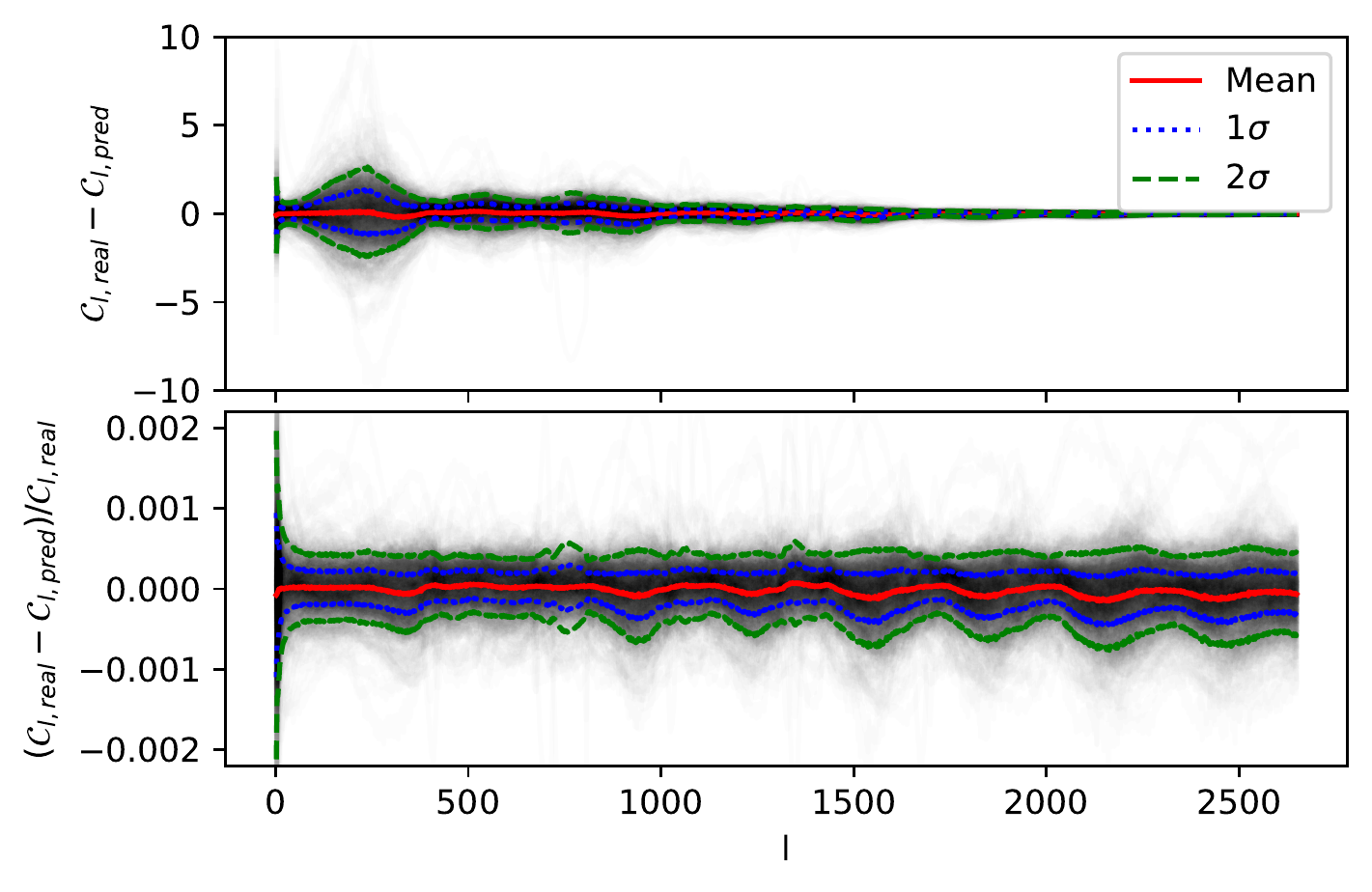}
    \caption{\textit{Top panel}: Difference between the real power spectra (computed with \texttt{CAMB}) and
    the predicted one (computed with the decoder model) for all the
    cosmological models of the test-set. \textit{Lower Panel}: Percentage difference between the real and the predicted power spectra for all the
    cosmological models of the test-set. 
    In red solid line are shown the mean values, while in green and blue dotted lines are shown the $1$ and $2$ standard deviation from the mean.}
    \label{fig:decoder}
\end{figure}

In the top panel of figure \ref{fig:decoder} we show
the difference between the predicted and the real power spectra
for all the cosmological parameters of the testing-set.
While in the
lower panel we show the percentage error in the prediction of each multipole.

The mean performance (shown in red line on figure \ref{fig:decoder}) is around $\approx 0.0018\%$ for all the multipoles.
We also plot as blue-dotted and green-dashed lines $1$ and $2$ standard deviation ($\sigma$).
As can be seen, the errors in all the multipoles are below $1\%$. 
This demonstrates the capacity of the decoder to compute the power spectrum from the cosmological parameters.

\begin{figure*}
    \includegraphics[width=1.1\linewidth, left]{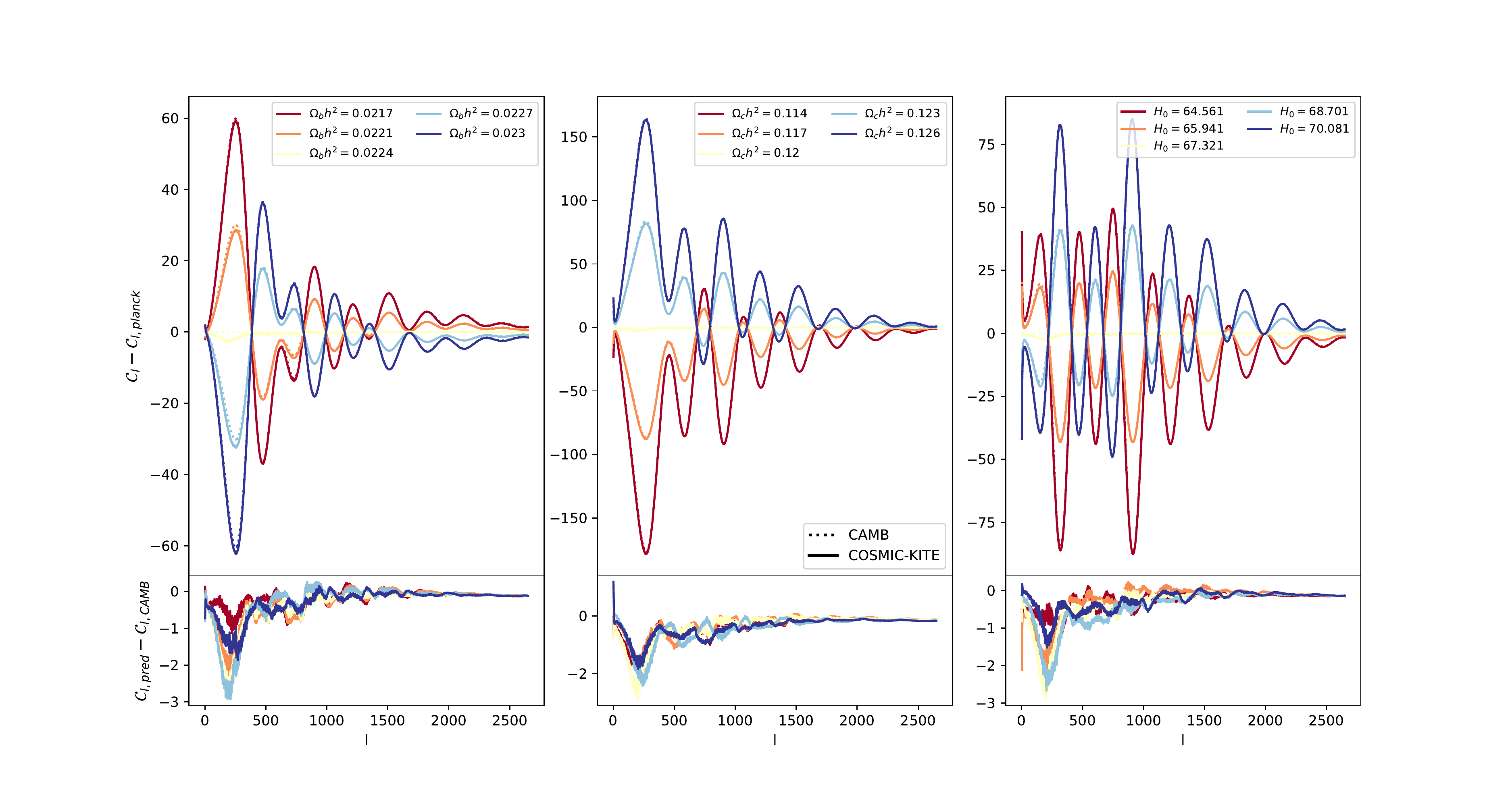}
    \includegraphics[width=1.1\linewidth, left]{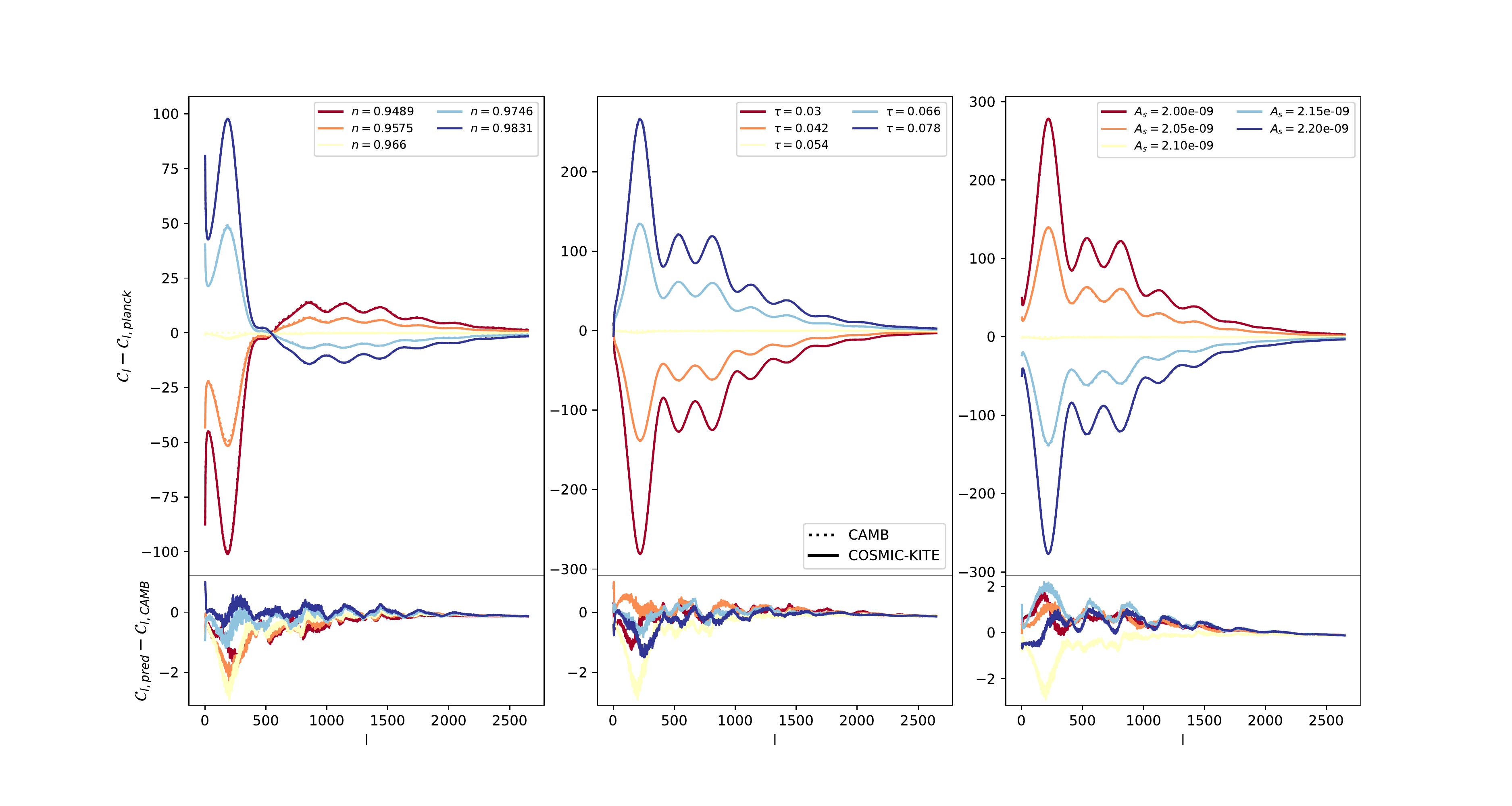}
    \caption{Analysis of the TT power spectra when varying one parameter at time while having all the rest fixed. 
    Each panel correspond to the variation of one cosmological parameter. Inside each panel, the upper plot shows the difference between the TT power spectrum
    with a fiducial planck cosmology and the TT spectrum with a cosmological parameter varying according to the color code in the
    right corner of the plot. For a better comparison we also add, in dotted lines, the power spectra spectra computed with \texttt{CAMB}.
    The lower plot shows the difference between the TT spectrum predicted with the decoder and the spectrum
    predicted with \texttt{CAMB}. 
    }
    \label{fig:1par_var}
\end{figure*}

Although the results shown in figure \ref{fig:decoder} demonstrate that the decoder can be used as forward model to compute the power
spectrum that correspond to a given set of cosmological parameters, it is interesting to
analyze the behaviour of such forward model when varying one parameter at time. 
In figure \ref{fig:1par_var} we show, in solid lines, the results of computing the power spectrum with the decoder varying one parameter at time while fixing all the other parameters to the fiducial planck values.
Each panel correspond to the variation of one cosmological parameter.
Inside each panel, the upper plot shows the absolute difference between the TT power spectrum with a fiducial planck cosmology and the TT spectrum with a cosmological parameter varying according to the color code in the right corner of the plot. For a better comparison we also add, in dotted lines, the power spectra spectra computed with 
\texttt{CAMB}.
The lower plot shows the absolute difference between the TT spectrum predicted with the decoder and the spectrum predicted with \texttt{CAMB}.

As it can be seen, the decoder recovered the expected trends when varying all the cosmological
parameters one by one.
These results demonstrate that the decoder is well trained and that it can be used as a forward model for a fully Bayesian analysis.

In general the computed theoretical power spectra are used to estimate a likelihood in a Bayesian analysis.
So, it is an interesting exercise to estimate the error introduced in a given likelihood when using the
decoder as a forward model for the estimation of the cosmological parameters.
In order to do that, we estimate a power spectrum for a fiducial Planck cosmology using the \texttt{CAMB} code. 
After that, we estimate the power spectra for random cosmologies using the \texttt{CAMB} code and the decoder as forward models.
Finally, following \cite{verde2008}, we compute the likelihood (eq. \ref{eq:lik}) between each random power spectrum and the fiducial one, and compare the results obtained using both forward models.

\begin{equation} \label{eq:lik}
\mathcal{L} = \sum_{l=50}^{2500} (2l+1)\left[ ln \left( \frac{C_{l}^{R}}{C_{l}^{P}} \right) + \frac{C_{l}^{P}}{C_{l}^{R}} - 1 \right]
\end{equation}

\begin{figure}
    \centering
    \includegraphics[scale=0.5]{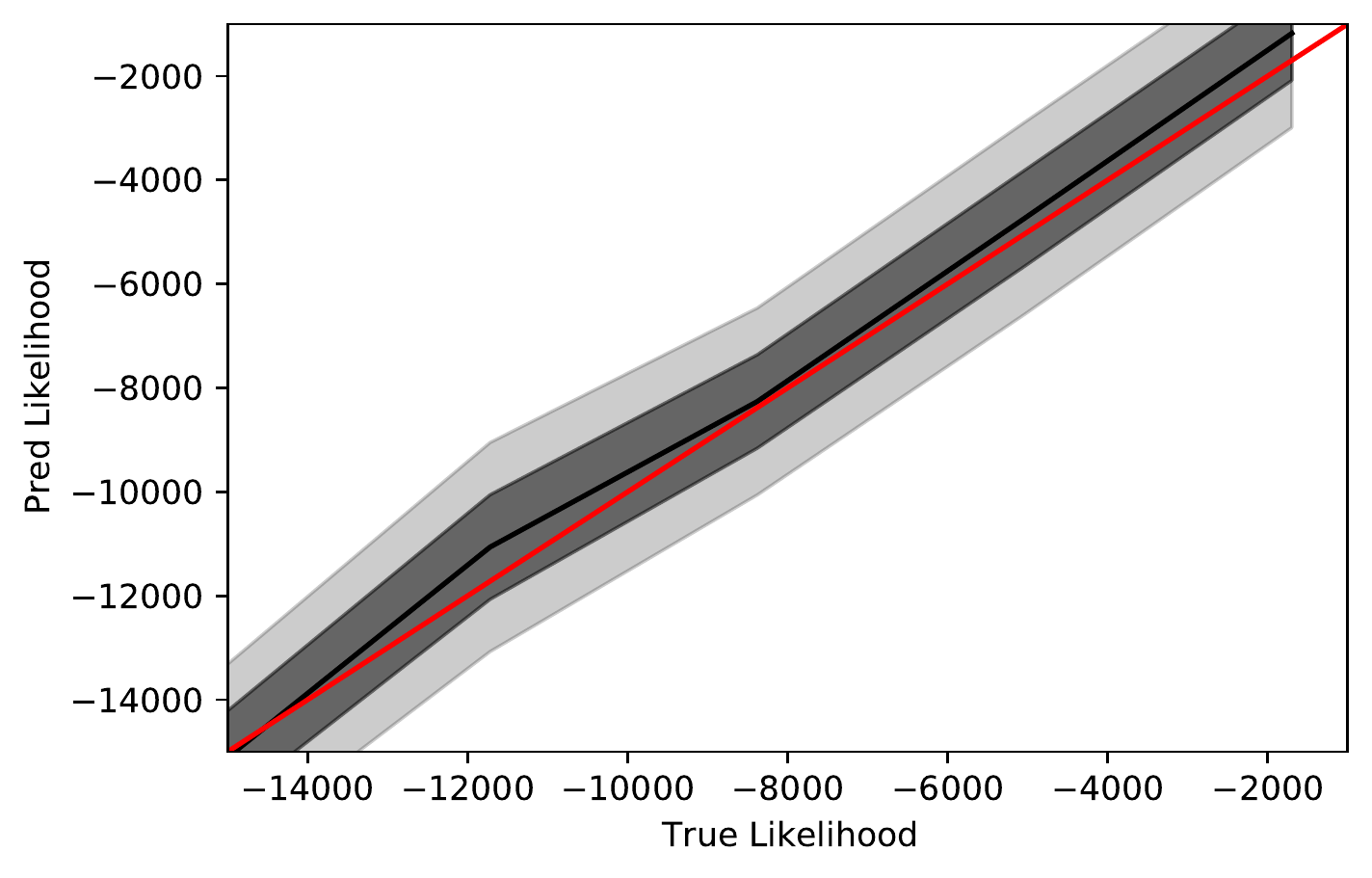}
    \caption{Comparison between the true likelihood estimated using \texttt{CAMB} as a forward model
    vs the predicted likelihood computed using the
    decoder as a forward model.}
    \label{fig:lik_comparison}
\end{figure}

In figure \ref{fig:lik_comparison} we show the predicted likelihood (computed with the decoder power spectra) vs the real likelihood (estimated with \texttt{CAMB} power spectra).

As it can be seen, the error introduced by the decoder in the prediction of the likelihood is very low in all the likelihood range.
This guarantee that if we perform a full Bayesian estimation
using the decoder as forward model we will obtain similar results as using the \texttt{CAMB} code.

\subsection{Estimating the full-posterior of cosmological parameters.}\label{sec:bayes}

Usually the cosmological parameters are estimated through a Bayesian analysis
where the theoretical power spectra enters inside a likelihood function.
In order to test if using the decoder as a forward model introduces any bias in the estimation of the
cosmological parameters, we performed a Bayesian analysis using Monte-Carlo Markov-Chains to predict the parameters of our fiducial Planck model.
For this analysis we use the Planck TT-LITE likelihood \citep{plik}, nevertheless we remark
that, as the decoder has a very good accuracy in predicting each multipole, any likelihood function can be used and the results will agree with the
one obtained using any other forward model.

\begin{figure*}
    \centering
    \includegraphics[scale=0.7]{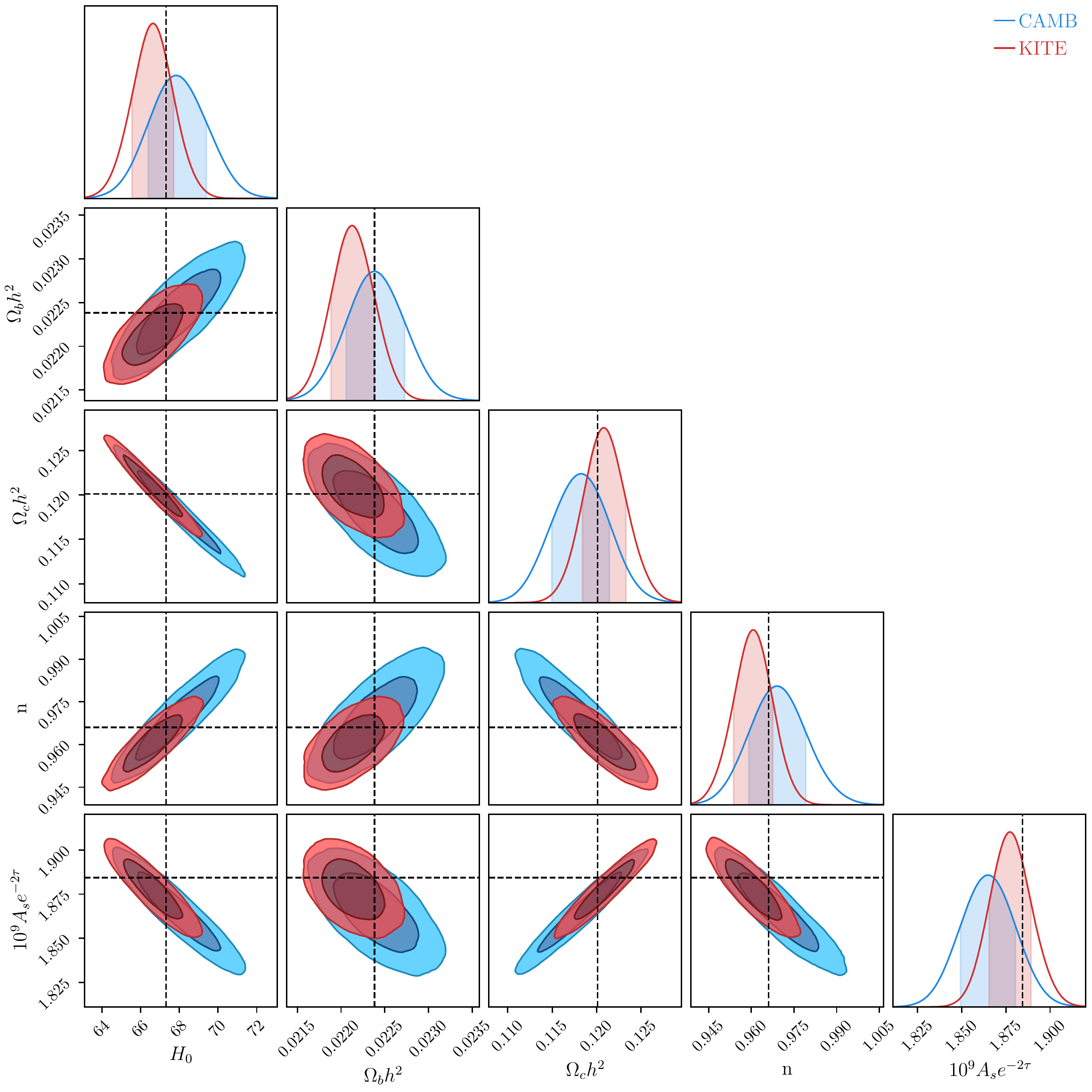}
    \caption{Full posterior contours obtained from an MCMC analysis with the Planck TT-LITE likelihood. In light blue are shown the results using the decoder as a forward model,
    while in red are shown the contours using the \texttt{CAMB} code as a forward model.}
    \label{fig:planck}
\end{figure*}

In figure \ref{fig:planck} we show the posterior contours for all the cosmological parameters.
In blue are shown the results using the decoder as forward model while in orange are shown the results using \texttt{CAMB} as forward model.
In addition we show in table \ref{tab:results_mcmc} the mean and marginal errors of the cosmological parameters found using \texttt{CAMB} and the decodes as forward models.

\begin{table}
    \centering
    \begin{tabular}{|c|c|c|}
        \hline
		Parameter & Decoder & \texttt{CAMB} \\
		\hline
		$H_{0}$ & $66.7^{+1.1}_{-1.1}$ & $67.8^{+1.6}_{-1.4}$ \\
		\hline
		$\Omega_{b}h^{2}$ & $0.02213^{+0.000025}_{-0.000024}$ & $0.2238^{+0.000035}_{-0.000035}$ \\
		\hline
	    $\Omega_{c}h^{2}$ & $0.1208^{+0.0025}_{-0.0024}$ & $0.1182\pm 0.0033$ \\
		\hline
		n & $0.9607^{+0.0069}_{-0.0069}$ & $0.9690^{+0.01}_{-0.0098}$ \\
		\hline
		$10^{9}A_{s}e^{-2\tau}$ & $1.877\pm 0.012$ & $1.865^{+0.016}_{-0.015}$ \\
		 
    \end{tabular}
    \caption{Mean cosmological parameters and marginal errors estimated with the MCMC analysis using the decoder and  \texttt{CAMB} as forward models. }
    \label{tab:results_mcmc}
\end{table}

It can be seen that, although there are some small deviations, the results using both forward models agree. 
This agreement comes from the fact that the decoder has a high accuracy in, predicting each multipole of the power spectra. 
While the small deviations comes from the fact that \texttt{COSMIC-KITE} is a fast approximated method, that do not predicted with perfect accuracy the power spectrum in the full multipole range.

It is worth remarking that although, for the sake of clarity we only show the estimated contours for a Planck-like cosmology, we tested the method for random cosmologies always finding a good agreement compared with the results obtained using \texttt{CAMB} as forward model.

\section{\texttt{COSMIC-KITE}: Examples of usage and profiling.} \label{sec:cosmickite}
In the previous sections we explicitly demonstrate the capacity of the encoder to predict the maximum-likelihood
cosmological parameters from a power spectrum and
of the decoder to estimate the TT power spectrum from
the cosmological parameters.
In this section we introduce a python software that uses the pre-trained encoder and decoder models to estimate the cosmological parameters and the power spectra respectively.

This software can be downloaded from \href{https://github.com/Martindelosrios/cosmic-kite}{https://github.com/Martindelosrios/cosmic-kite}
or directly installed in a python environment with:

\begin{verbatim}
 pip install 
 git+https://github.com/Martindelosrios/cosmic-kite
\end{verbatim}

The two main functions of the software are \texttt{pars2ps} and \texttt{ps2pars} that computes
the power spectrum from the cosmological parameters (using the decoder model) and the cosmological parameters from the
power spectrum (using the encoder model).

At the github repository we provided examples of basic usage of both functions, as well as a
notebook\footnote{\href{https://github.com/Martindelosrios/cosmic-kite/blob/main/Examples/Figures.ipynb}{https://github.com/Martindelosrios/cosmic-kite/blob/main/Examples/Figures.ipynb}} that can be run to produce the figures
presented in this paper.

As explained before, one of the main advantages of using \texttt{cosmic-kite} as forward model is that is much
faster than codes as \texttt{CAMB}.

In order to quantify the velocity of both softwares
we compute the power spectra for $1000$ random cosmologies one by one.
\texttt{CAMB} took $\approx 6900$ seconds while
\texttt{cosmic-kite} took $\approx 150 $ seconds to
compute the power spectra for the random cosmologies in a laptop with 8 cores, representing an increase
on the velocity of $\approx 45x$.
In addition, \texttt{cosmic-kite} has the option of computing all the power spectra of a set of
cosmological models in one shot. \texttt{cosmic-kite} took $\approx 0.82$ seconds to compute
all the power spectra in this mode, representing
an increase on the velocity of more than $\approx 8400x$.

\section{Discussion and Conclusions.} \label{sec:conclusions}

In this work, we present the results of the study of the CMB TT power spectrum through an auto-encoder in which the latent variables correspond to the cosmological parameters.
Auto-encoders are neural network with an specific architecture that in the middle has a hidden layer with few neurons that is called the bottle-neck layer.
In this kind of model the input and the output are the same, and, in general, has a much
higher dimension than the bottle-neck layer. 
So, when training the method, the auto-encoder
is forced to encode the input into the latent variables and then decode them to reconstruct the input.
In this implementation we introduce as input/output the CMB TT power spectra and force it to be encoded into $6$ latent variables.
Additionally, we add a term to the loss function that force the latent variables to be the cosmological parameters.
Our method was calibrated with $80000$ power spectra computed with the code \texttt{CAMB} from random cosmological parameters.
The advantage of our method is two-fold. On one hand
the encoder can be used as a fast way of estimating the maximum
likelihood parameters from a given power spectrum. 
On the other hand, the decoder can be used to
compute the power spectrum from the cosmological parameters.
We show that the encoder achieves a high precision on predicting all the cosmological parameters.
We also demonstrate that the decoder is capable of estimate the power spectrum from a given set of
cosmological parameters and obtained similar results as the one obtained by the code \texttt{CAMB}.
In addition we explicitly show that the decoder is capable to obtained the expected trends when varying one
parameter at a time and that can be used as a forward model in a Bayesian analysis without introducing any bias on the estimated parameters.

We developed a python software that uses the pre-trained encoder and decoder models to estimate the cosmological parameters and the power spectra respectively.
This software can be downloaded from \href{https://github.com/Martindelosrios/cosmic-kite}{https://github.com/Martindelosrios/cosmic-kite}
or directly installed in a python environment with the \texttt{pip} command.

Although this algorithm does not improve the precision of the measurements compared with the traditional methods, it reduces
significantly the computation time and represents the first attempt towards forcing the latent variables to have a physical interpretation.
 
\section*{Acknowledgements}

I thank Rogerio Rosenfeld, Bruno Sanch\'ez, Juan B. Cabral and Mariano Dom\'inguez for all the helpful discussions.
I speccially thank to Antonino Troja for his help.

This research was partially supported by the Laborat\'orio Interinstitucional de e-Astronomia (LIneA),
the Sao Paulo State Research Agency (FAPESP) through grant
2019/08852-2, the Spanish Research Agency (Agencia
Estatal de Investigación) through the Grant IFT Centro de Excelencia Severo
Ochoa No CEX2020-001007-S, funded by MCIN/AEI/10.13039/501100011033 and by 'Comunidad
Aut\'onoma de Madrid' through the grant SI2/PBG/2020-00005. 

\section*{DATA AVAILABILITY}

The data underlying this article are available
in \href{https://github.com/Martindelosrios/cosmic-kite}{https://github.com/Martindelosrios/cosmic-kite}. 

\bibliographystyle{mn2e}
\bibliography{references}

\begin{thebibliography}{}

\bibitem[\protect\citeauthoryear{{Aghanim}, {Majumdar} \& {Silk}}{{Aghanim}
  et~al.}{2008}]{secondaries_ans}
{Aghanim} N.,  {Majumdar} S.,    {Silk} J.,  2008, Reports on Progress in
  Physics, 71, 066902

\bibitem[\protect\citeauthoryear{{Alonso}, {Sanchez}, {Slosar} \& {LSST Dark
  Energy Science Collaboration}}{{Alonso} et~al.}{2019}]{namaster}
{Alonso} D.,  {Sanchez} J.,  {Slosar} A.,    {LSST Dark Energy Science
  Collaboration} 2019, \mnras, 484, 4127

\bibitem[\protect\citeauthoryear{{Auld}, {Bridges} \& {Hobson}}{{Auld}
  et~al.}{2008}]{cosmonet2}
{Auld} T.,  {Bridges} M.,    {Hobson} M.~P.,  2008, \mnras, 387, 1575

\bibitem[\protect\citeauthoryear{{Auld}, {Bridges}, {Hobson} \& {Gull}}{{Auld}
  et~al.}{2007}]{cosmonet1}
{Auld} T.,  {Bridges} M.,  {Hobson} M.~P.,    {Gull} S.~F.,  2007, \mnras, 376,
  L11

\bibitem[\protect\citeauthoryear{{Blas}, {Lesgourgues} \& {Tram}}{{Blas}
  et~al.}{2011}]{class2}
{Blas} D.,  {Lesgourgues} J.,    {Tram} T.,  2011, \jcap, 2011, 034

\bibitem[\protect\citeauthoryear{Chollet et~al.,}{Chollet
  et~al.}{2015}]{keras}
Chollet F.,  et~al.,, 2015, Keras

\bibitem[\protect\citeauthoryear{Developers}{Developers}{2021}]{tensorflow}
Developers T., , 2021, TensorFlow

\bibitem[\protect\citeauthoryear{Dodelson}{Dodelson}{2003}]{dodelson}
Dodelson S.,  2003, Modern cosmology.
Academic Press

\bibitem[\protect\citeauthoryear{{Fendt} \& {Wandelt}}{{Fendt} \&
  {Wandelt}}{2007}]{pico}
{Fendt} W.~A.,  {Wandelt} B.~D.,  2007, \apj, 654, 2

\bibitem[\protect\citeauthoryear{G{\'o}rski, {Hivon}, {Banday}, {Wandelt},
  {Hansen}, {Reinecke} \& {Bartelmann}}{G{\'o}rski et~al.}{2005}]{healpix}
G{\'o}rski K.~M.,  {Hivon} E.,  {Banday} A.~J.,  {Wandelt} B.~D.,  {Hansen}
  F.~K.,  {Reinecke} M.,    {Bartelmann} M.,  2005, \apj, 622, 759

\bibitem[\protect\citeauthoryear{Hinton \& Salakhutdinov}{Hinton \&
  Salakhutdinov}{2006}]{autoencoders}
Hinton G.~E.,  Salakhutdinov R.~R.,  2006, Science, 313, 504

\bibitem[\protect\citeauthoryear{{Hivon}, {G{\'o}rski}, {Netterfield}, {Crill},
  {Prunet} \& {Hansen}}{{Hivon} et~al.}{2002}]{hivon}
{Hivon} E.,  {G{\'o}rski} K.~M.,  {Netterfield} C.~B.,  {Crill} B.~P.,
  {Prunet} S.,    {Hansen} F.,  2002, \apj, 567, 2

\bibitem[\protect\citeauthoryear{{Khatri}}{{Khatri}}{2019}]{khatri}
{Khatri} R.,  2019, \jcap, 2019, 039

\bibitem[\protect\citeauthoryear{Kingma \& Ba}{Kingma \& Ba}{2017}]{adam}
Kingma D.~P.,  Ba J., , 2017, Adam: A Method for Stochastic Optimization

\bibitem[\protect\citeauthoryear{{Lesgourgues}}{{Lesgourgues}}{2011}]{class1}
{Lesgourgues} J.,  2011, arXiv e-prints, p. arXiv:1104.2932

\bibitem[\protect\citeauthoryear{Lewis, Challinor \& Lasenby}{Lewis
  et~al.}{2000}]{camb}
Lewis A.,  Challinor A.,    Lasenby A.,  2000, \apj, 538, 473

\bibitem[\protect\citeauthoryear{{Montefalcone}, {Abitbol}, {Kodwani} \&
  {Grumitt}}{{Montefalcone} et~al.}{2020}]{montefalcone}
{Montefalcone} G.,  {Abitbol} M.~H.,  {Kodwani} D.,    {Grumitt} R.~D.~P.,
  2020, arXiv e-prints, p. arXiv:2011.01433

\bibitem[\protect\citeauthoryear{{N{\o}rgaard-Nielsen}}{{N{\o}rgaard-Nielsen}}{2012}]{norgaard}
{N{\o}rgaard-Nielsen} H.~U.,  2012, \apss, 340, 161

\bibitem[\protect\citeauthoryear{Pedregosa, Varoquaux, Gramfort, Michel,
  Thirion, Grisel, Blondel, Prettenhofer, Weiss, Dubourg, Vanderplas, Passos,
  Cournapeau, Brucher, Perrot \& Duchesnay}{Pedregosa
  et~al.}{2011}]{scikit-learn}
Pedregosa F.,  Varoquaux G.,  Gramfort A.,  Michel V.,  Thirion B.,  Grisel O.,
   Blondel M.,  Prettenhofer P.,  Weiss R.,  Dubourg V.,  Vanderplas J.,
  Passos A.,  Cournapeau D.,  Brucher M.,  Perrot M.,    Duchesnay E.,  2011,
  Journal of Machine Learning Research, 12, 2825

\bibitem[\protect\citeauthoryear{{Perraudin}, {Defferrard}, {Kacprzak} \&
  {Sgier}}{{Perraudin} et~al.}{2019}]{deepsphere}
{Perraudin} N.,  {Defferrard} M.,  {Kacprzak} T.,    {Sgier} R.,  2019,
  Astronomy and Computing, 27, 130

\bibitem[\protect\citeauthoryear{{Planck Collaboration}, {Aghanim}, {Akrami},
  {Ashdown}, {Aumont} \& {Baccigalupi}}{{Planck Collaboration}
  et~al.}{2020}]{planck2018}
{Planck Collaboration} {Aghanim} N.,  {Akrami} Y.,  {Ashdown} M.,  {Aumont} J.,
     {Baccigalupi} C.,  2020, \aap, 641, A6

\bibitem[\protect\citeauthoryear{{Planck Collaboration}, {Aghanim}, {Akrami},
  {Ashdown}, {Aumont}, {Baccigalupi}, {Ballardini}, {Banday}, {Barreiro},
  {Bartolo}, {Basak}, {Benabed}, {Bernard}, {Bersanelli} \& et al.}{{Planck
  Collaboration} et~al.}{2020}]{plik}
{Planck Collaboration} {Aghanim} N.,  {Akrami} Y.,  {Ashdown} M.,  {Aumont} J.,
   {Baccigalupi} C.,  {Ballardini} M.,  {Banday} A.,  {Barreiro} R.,  {Bartolo}
  N.,  {Basak} S.,  {Benabed} K.,  {Bernard} J.~P.,  {Bersanelli} M.,    et al.
  2020, \aap, 641, A5

\bibitem[\protect\citeauthoryear{{Riess}, {Macri}, {Hoffmann}, {Scolnic},
  {Casertano}, {Filippenko}, {Tucker}, {Reid}, {Jones}, {Silverman},
  {Chornock}, {Challis}, {Yuan}, {Brown} \& {Foley}}{{Riess}
  et~al.}{2016}]{riess}
{Riess} A.~G.,  {Macri} L.~M.,  {Hoffmann} S.~L.,  {Scolnic} D.,  {Casertano}
  S.,  {Filippenko} A.~V.,  {Tucker} B.~E.,  {Reid} M.~J.,  {Jones} D.~O.,
  {Silverman} J.~M.,  {Chornock} R.,  {Challis} P.,  {Yuan} W.,  {Brown} P.~J.,
     {Foley} R.~J.,  2016, ArXiv e-prints

\bibitem[\protect\citeauthoryear{Rumelhart, Hinton \& Williams}{Rumelhart
  et~al.}{1986}]{autoencoder}
Rumelhart D.~E.,  Hinton G.~E.,    Williams R.~J.,  1986, Learning Internal
  Representations by Error Propagation.
MIT Press, Cambridge, MA, USA, p. 318–362

\bibitem[\protect\citeauthoryear{{Spurio Mancini}, {Piras}, {Alsing},
  {Joachimi} \& {Hobson}}{{Spurio Mancini} et~al.}{2021}]{cosmopower}
{Spurio Mancini} A.,  {Piras} D.,  {Alsing} J.,  {Joachimi} B.,    {Hobson}
  M.~P.,  2021, arXiv e-prints, p. arXiv:2106.03846

\bibitem[\protect\citeauthoryear{Verde}{Verde}{2008}]{verde2008}
Verde L., , 2008, A practical guide to Basic Statistical Techniques for Data
  Analysis in Cosmology

\end{thebibliography}


\appendix
\section{Auto-encoding the TE and EE power spectra.}\label{ap:TE-EE}

Although the main results of the paper were obtained using only the TT power spectrum, the architecture and the
idea of forcing the latent variables to have a physical
interpretation is completely agnostic to
the dataset and can be easily extended to any other physical problem using the proper training set.

In this appendix we show that the same formalism can be used with the EE and TE power spectra.

In order to do this, we generate a dataset computing the TE and EE power spectra for the same cosmologies of the dataset presented on section \ref{sec:set_datos}.

Using the same architecture presented on section \ref{sec:metodos_cmb}
we trained two new auto-encoder. One for the TE power spectrum and one for the EE power spectrum.

\begin{figure}
    \centering
    \includegraphics[scale=0.6]{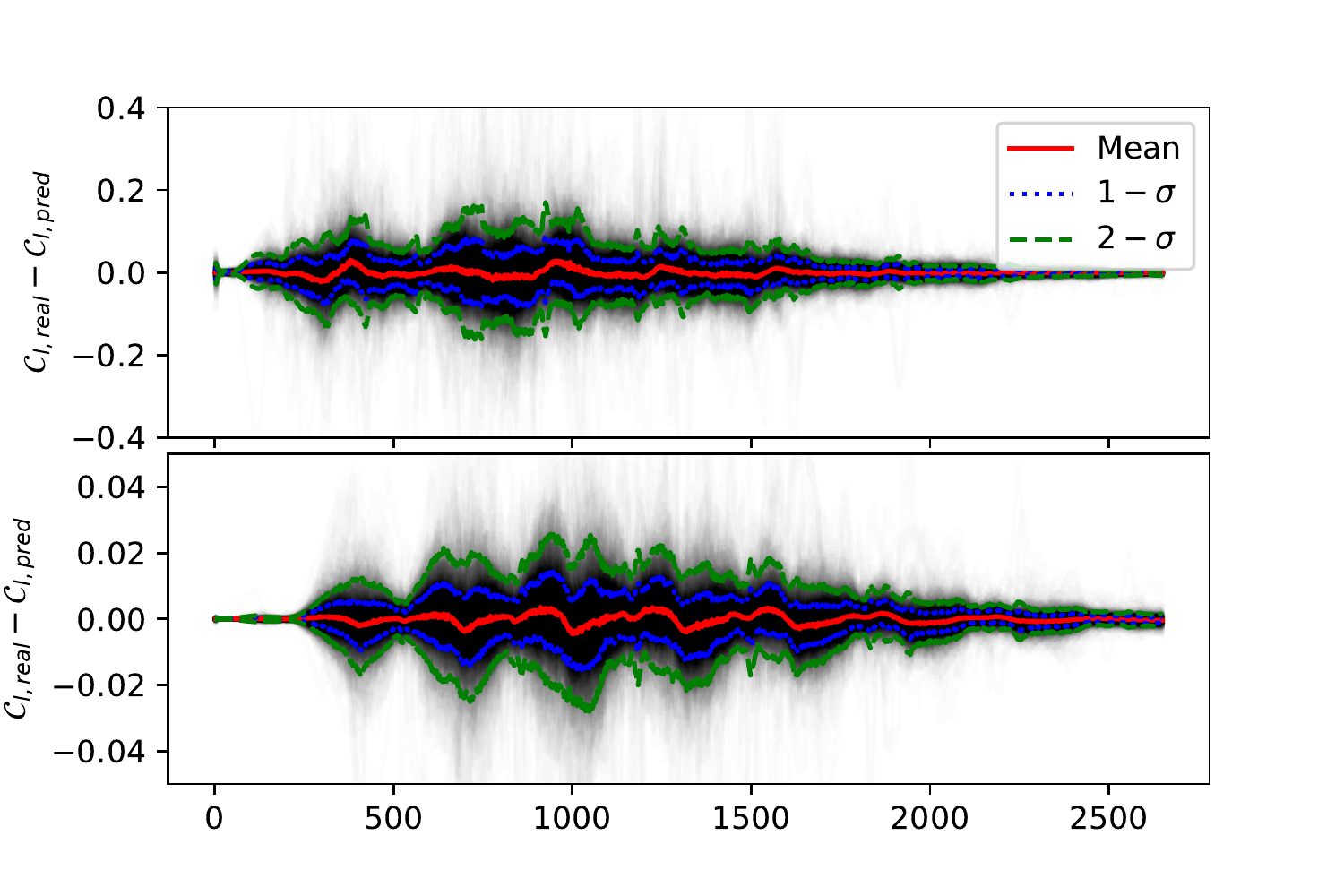}
    \caption{\textit{Top panel}: Difference between the real TE power spectra (computed with \texttt{CAMB}) and
    the predicted one (computed with the decoder model) for all the cosmological models of the test-set. \textit{Lower Panel}: Difference between the real EE power spectra (computed with \texttt{CAMB}) and
    the predicted one (computed with the decoder model) for all the cosmological models of the test-set.
    In red solid line are shown the mean values, while in green and blue dotted lines are shown the $1$ and $2$ standard deviation from the mean.}
    \label{fig:TE-EE}
\end{figure}

In figure \ref{fig:TE-EE} we show the comparison between the predicted and the real (computed with \texttt{CAMB}) TE and EE power spectra for the cosmologies of the test-set.
It can be seen that both auto-encoders predict with high precision the corresponding power spectra.

\begin{figure*}
    \includegraphics[width=1.1\linewidth, left]{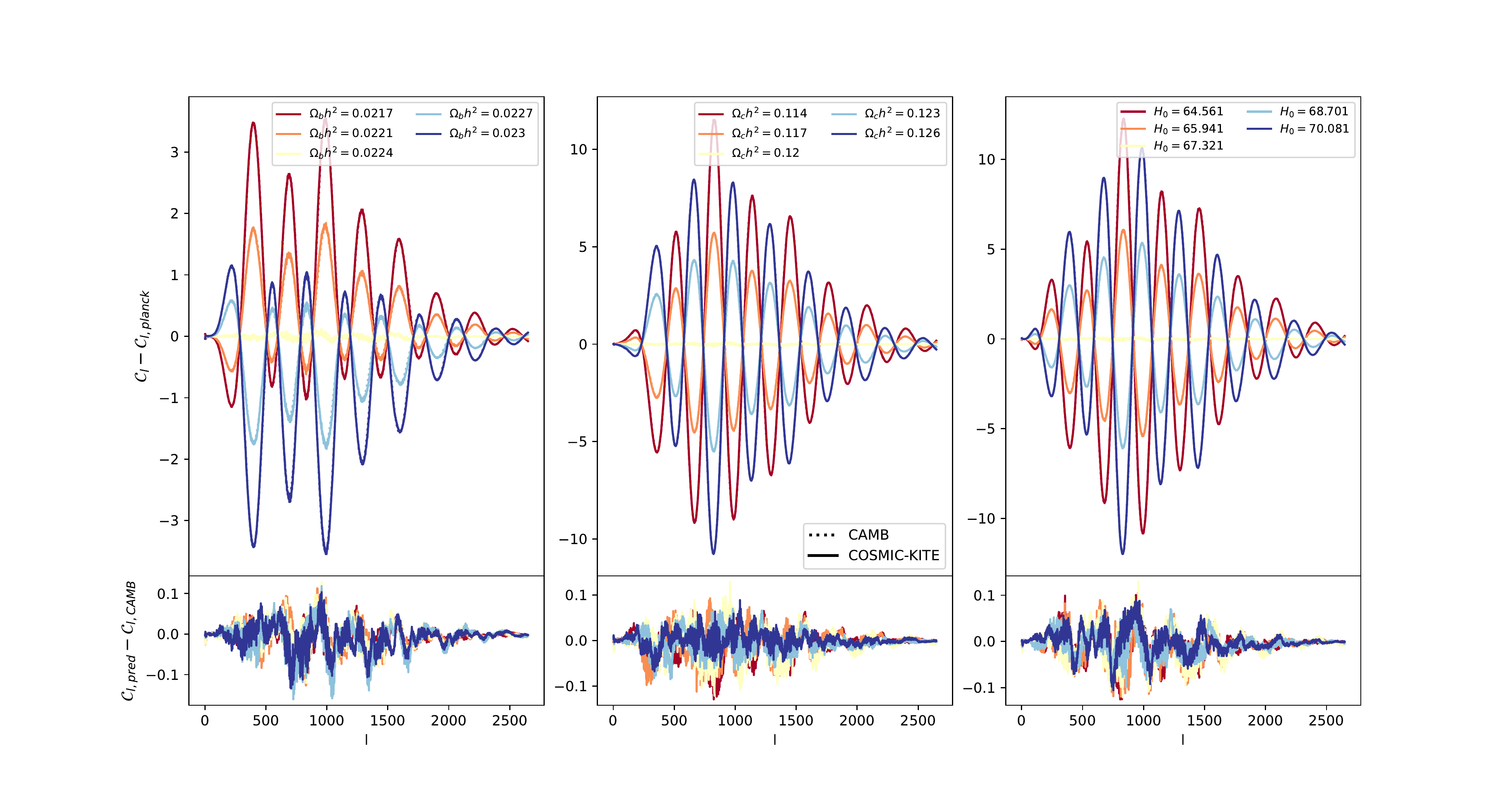}
    \includegraphics[width=1.1\linewidth, left]{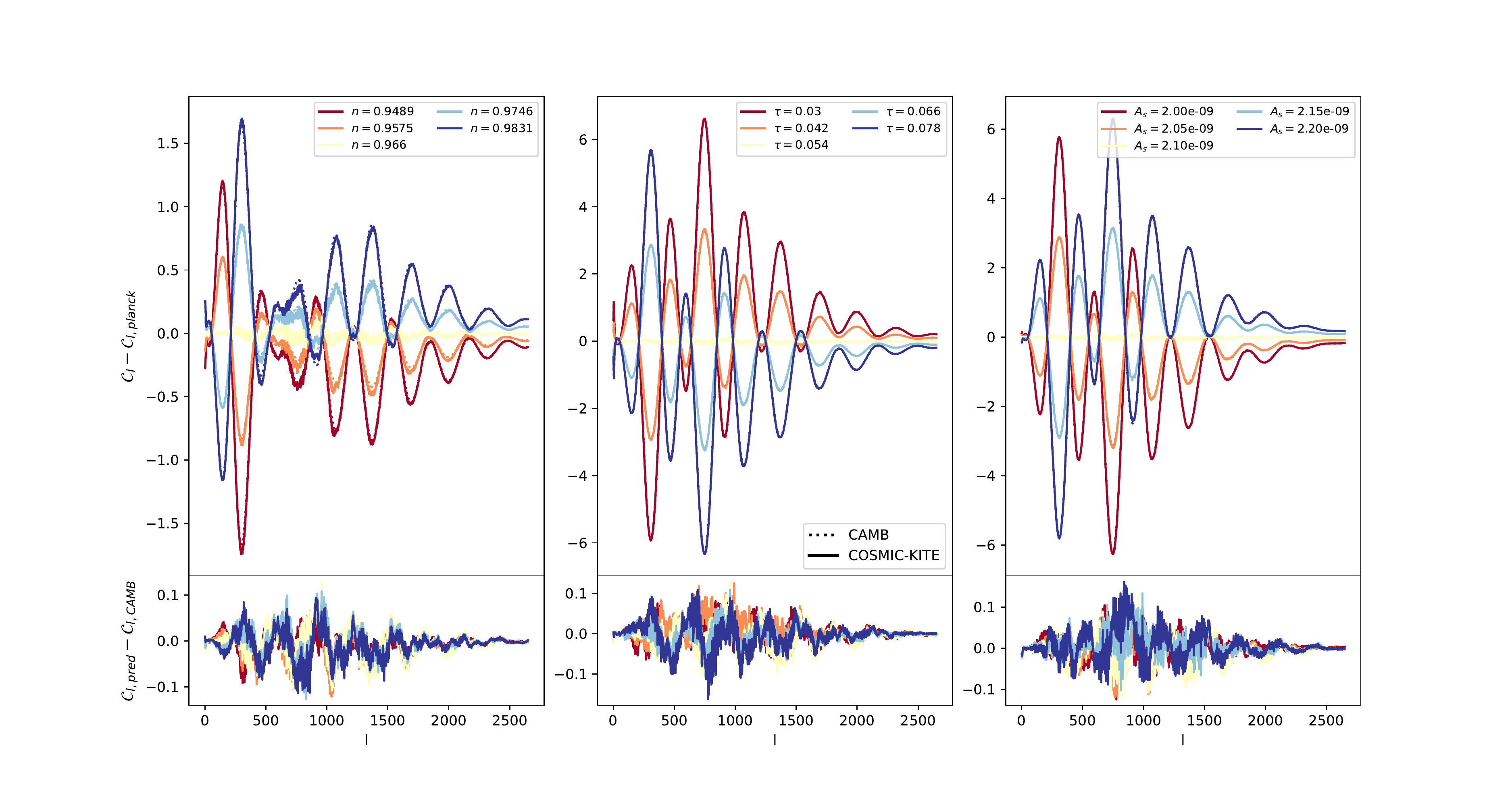}
    \caption{Analysis of the TE power spectra when varying one parameter at time while having all the rest fixed.     
    Each panel correspond to the variation of one cosmological parameter. Inside each panel, the upper plot shows the difference between the TE power spectrum
    with a fiducial planck cosmology and the TE spectrum with a cosmological parameter varying according to the color code in the
    right corner of the plot. For a better comparison we also add, in dotted lines,the power spectra spectra computed with \texttt{CAMB}.
    The lower plot shows the difference between the TE spectrum predicted with the decoder and the spectrum
    predicted with \texttt{CAMB}. 
    }
    \label{fig:TE}
\end{figure*}

\begin{figure*}
    \includegraphics[width=1.1\linewidth, left]{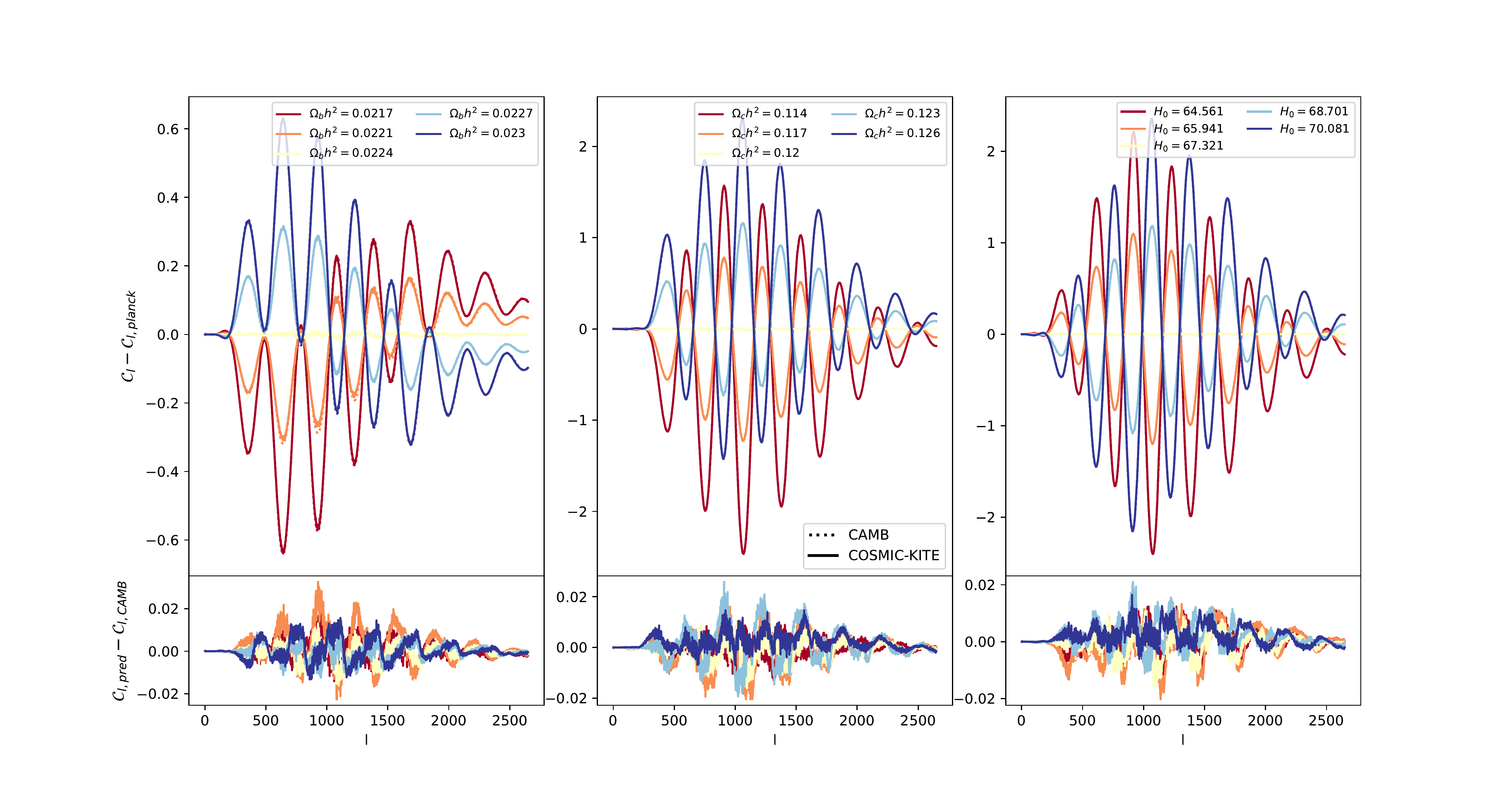}
    \includegraphics[width=1.1\linewidth, left]{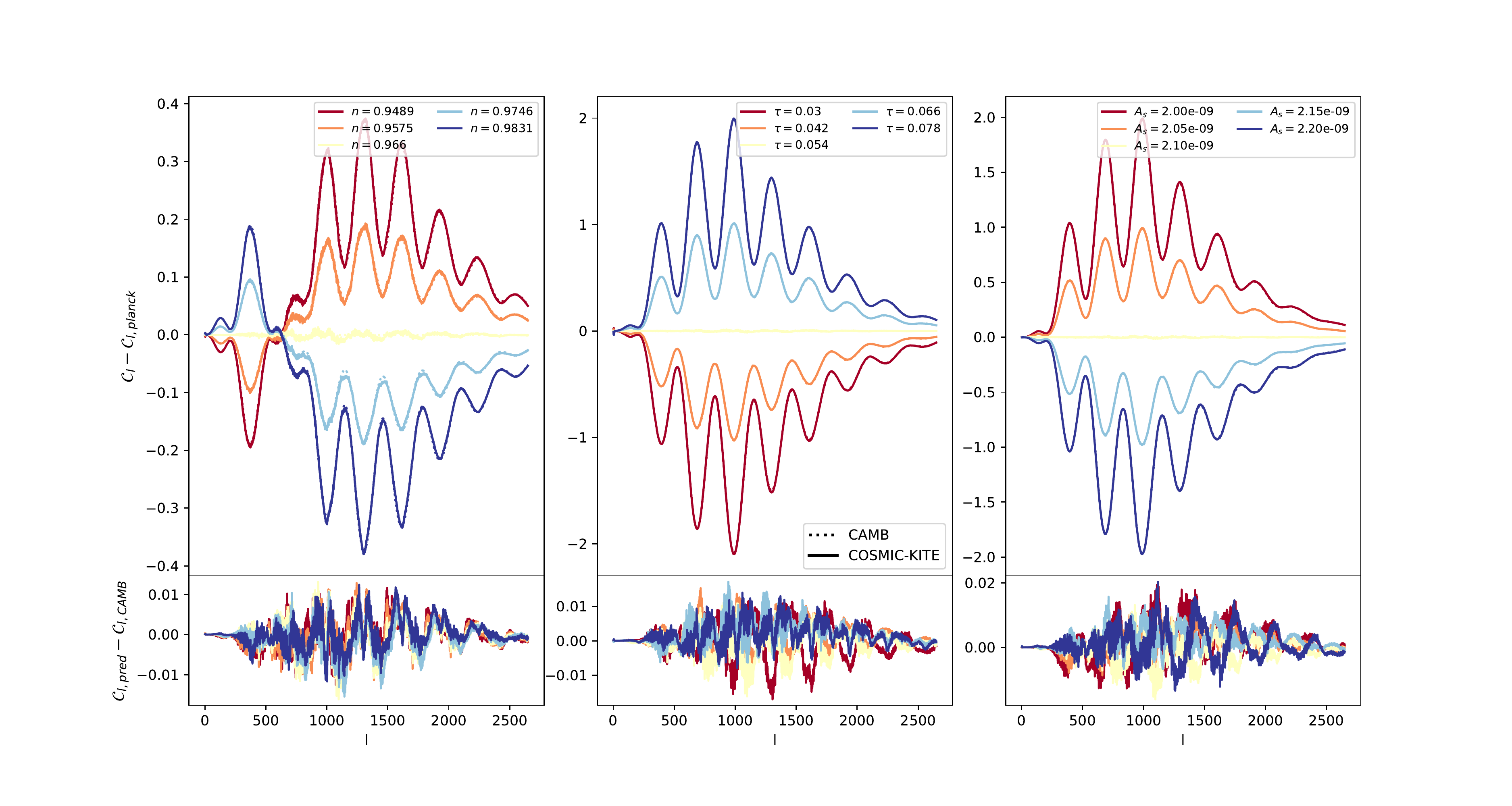}
    \caption{Analysis of the EE power spectra when varying one parameter at time having while all the rest fixed.  
    Each panel correspond to the variation of one cosmological parameter. Inside each panel, the upper plot shows the difference between the EE power spectrum
    with a fiducial planck cosmology and the EE spectrum with a cosmological parameter varying according to the color code in the
    right corner of the plot. For a better comparison we also add, in dotted lines,the power spectra spectra computed with \texttt{CAMB}.
    The lower plot shows the difference between the EE spectrum predicted with the decoder and the spectrum
    predicted with \texttt{CAMB}. 
    }
    \label{fig:EE}
\end{figure*}

In addition we show in figures \ref{fig:EE} and \ref{fig:TE}
the errors of the TE and EE power spectra when varying one parameter at time and having all the rest fixed.

It can be seen that the auto-encoders predict with high accuracy all the power spectra in the full multipole range and in the full cosmological parameter space of interest.
This demonstrate that the \texttt{COSMIC-KITE} code can be used as forward model in any MCMC analysis.

\label{lastpage}

\end{document}